\documentclass{aa}  
\usepackage[varg]{txfonts}
\usepackage{ctable}
\usepackage{graphicx}
\usepackage{txfonts}
\usepackage{hyperref}
\usepackage{amsmath}
\usepackage{ulem}
\usepackage{natbib}
\bibpunct{(}{)}{;}{a}{}{,}

\newcommand{\nfrb}{14}
\newcommand{\rmobs}{\ensuremath{\mathrm{RM_{obs}}}}
\newcommand{\rmmodel}{\ensuremath{\mathrm{RM_{model}}}}
\newcommand{\rmmw}{\ensuremath{\mathrm{RM_{MW}}}}
\newcommand{\rmigm}{\ensuremath{\mathrm{RM_{IGM}}}}
\newcommand{\rmhalo}{\ensuremath{\mathrm{RM_{fg}^{halo}}}}
\newcommand{\rmhaloij}{\ensuremath{\mathrm{RM}_{\mathrm{fg},~ij}^{\mathrm{halo}} } }

\newcommand{\rmhost}{\ensuremath{\mathrm{RM_{host}}}}
\newcommand{\rmhhost}{\ensuremath{\mathrm{RM_{host}^{halo}}}}
\newcommand{\rmlhost}{\ensuremath{\mathrm{RM_{host}^{local}}}}
\newcommand{\rmlocal}{\ensuremath{\mathrm{RM_{host}^{local}}}}
\newcommand{\rmeg}{\ensuremath{\mathrm{RM_{eg}}}}
\newcommand{\rmunits}{\ensuremath{\rm rad \, m^{-2}}}
\newcommand{\alphafh}{ \ensuremath{\cos \alpha_{\rm fg}}}
\newcommand{\alphahh}{ \ensuremath{\cos \alpha_{\rm hh}}}
\newcommand{\alphahl}{ \ensuremath{\cos \alpha_{\rm hl}}}

\newcommand{\dmhostdir}{\ensuremath{\mathrm{DM_{host}^{direct}}}}
\newcommand{\dmhostism}{\ensuremath{\mathrm{DM_{host}^{ism}}}}
\newcommand{\dmlocal}{\ensuremath{\mathrm{DM_{host}^{local}}}}

\newcommand{\dmfrb}{\ensuremath{\mathrm{DM_{obs}}}}

\newcommand{\fgas}{\ensuremath{f_\mathrm{gas}}}
\newcommand{\bhalos}{\ensuremath{B_\mathrm{fg}^{\mathrm{halo}}}}
\newcommand{\bhosts}{\ensuremath{B_\mathrm{host}^{\mathrm{halo}}}}
\newcommand{\blocal}{\ensuremath{B_\mathrm{host}^{\mathrm{local}}}}

\newcommand{\fdiff}{\ensuremath{f_\mathrm{d}}}
\newcommand{\fdiffz}{\ensuremath{f_\mathrm{d}\left(z\right)}}
\newcommand{\figm}{\ensuremath{f_\mathrm{igm}}}
\newcommand{\ficm}{\ensuremath{f_\mathrm{icm}}}
\newcommand{\fcgm}{\ensuremath{f_\mathrm{cgm}}}

\newcommand{\bmag}{\ensuremath{B_{\parallel}}}

\begin{document} 

\title{Magnetic fields in galactic environments probed by fast radio bursts}

\author{Ilya S. Khrykin\inst{1}
        \and 
        Nicolas Tejos\inst{1}
        \and 
        J. Xavier Prochaska\inst{2,3,4,5}
        \and 
        Alexandra Mannings\inst{2}
        \and 
        Lluis Mas-Ribas\inst{2} 
        \and 
        Kentaro Nagamine\inst{3,7,8,9,10}
        \and 
        Khee-Gan Lee\inst{3,6}
        \and
        Bryan~M. Gaensler\inst{2,11,12}
        \and
        Zhao Joseph Zhang\inst{7}
        \and
        Lucas Bernales--Cortes\inst{1}
        }

\institute{Instituto de F\'isica, Pontificia   Universidad Cat\'olica de Valpara\'iso, Casilla 4059, Valpara\'iso, Chile\\
   \email{i.khrykin@gmail.com}
   \and
    Department of Astronomy and Astrophysics, University of California, Santa Cruz, 1156 High Street, Santa Cruz, CA 95064, USA
    \and
    Kavli IPMU (WPI), UTIAS, The University of Tokyo, Kashiwa, Chiba 277-8583, Japan
    \and
    Division of Science, National Astronomical Observatory of Japan, 2-21-1 Osawa, Mitaka, Tokyo 181-8588, Japan
    \and
    Simons Pivot Fellow
    \and
    Center for Data-Driven Discovery, Kavli IPMU (WPI), UTIAS, The University of Tokyo, Kashiwa, Chiba 277-8583, Japan
    \and 
    Theoretical Astrophysics, Department of Earth \& Space Science, Graduate School of Science, The University of Osaka, 1-1 Machikaneyama, Toyonaka, Osaka 560-0043, Japan
    \and
    Theoretical Joint Research, Forefront Research Center, Graduate School of Science, The University of Osaka, 1-1 Machikaneyama, Toyonaka, Osaka 560-0043, Japan
    \and
    Department of Physics \& Astronomy, University of Nevada, Las Vegas, 4505 S. Maryland Pkwy, Las Vegas, NV 89154-4002, USA
    \and
    Nevada Center for Astrophysics, University of Nevada, Las Vegas, 4505 S. Maryland Pkwy, Las Vegas, NV 89154-4002, USA
    \and
    Dunlap Institute for Astronomy and Astrophysics, University of Toronto, 50 St. George Street, Toronto, ON M5S 3H4, Canada
    \and
    David A.\ Dunlap Department of Astronomy and Astrophysics, University of Toronto, 50 St. George Street, Toronto, ON M5S 3H4, Canada}

\date{\today}
 
\abstract
   {Fast radio bursts (FRBs) are extragalactic, bright, millisecond radio pulses emitted by unknown sources. FRBs constitute a unique probe of various astrophysical and cosmological environments via their characteristic dispersion (DM) and Faraday rotation (RM) measures that encode information about the ionised gas traversed by the radio waves along the FRB line of sight. In this work, we analysed the observed RM measured for \nfrb~localised FRBs in the $0.05 \lesssim z_{\rm frb} \lesssim 0.5$ redshift range, in order to infer the total magnetic field, $B$, in various galactic environments. Additionally, we calculated \fgas~- the average fraction of baryons in the ionised CGM.
   We built a spectroscopic dataset of FRB foreground galaxy halos, acquired with VLT/MUSE observations 
   and by the FLIMFLAM collaboration. We developed 
   a novel Bayesian statistical algorithm and used it to
   correlate information on the individual intervening halos with the observed \rmobs. This approach allowed us to disentangle the magnetic fields present in various environments traversed by the FRB sight lines. Our analysis yields the first direct FRB constraints on the strength of magnetic fields in the interstellar medium (ISM) (\blocal) and in the halos (\bhosts) of FRB host galaxies, as well as in the halos of foreground galaxies and groups (\bhalos). Assuming no field reversals, we find that the average magnetic field strength in the ISM of the FRB host galaxies is $\blocal = 5.4^{+1.1}_{-0.9}~\mu{\rm G}$. Additionally, we placed an upper limit on the average magnetic field strength in FRB host halos, $\bhosts \lesssim 4.8~\mu{\rm G}$, and in foreground intervening halos, $\bhalos \lesssim 4.3~\mu{\rm G}$. Moreover, we estimated the average fraction of cosmic baryons inside $10 \lesssim \log_{10} \left( M_{\rm halo} / M_{\odot}\right) \lesssim 13.1$ halos to be $\fgas = 0.45^{+0.21}_{-0.19}$.
   We find that the magnetic field strengths inferred in this work are in good agreement with previous measurements. In contrast to previous studies that analysed FRB RMs and have not considered contributions from the halos of the foreground and/or FRB host galaxies, we show that halos can contribute a non-negligible amount of RM and must be taken into account when analysing future FRB samples.}

   \keywords{Galaxies: halos -- Galaxies: magnetic fields -- ISM: magnetic fields}

   \maketitle

\section{Introduction}
\label{sec:intro}

Over the last decades,  extragalactic millisecond radio transients, termed fast radio bursts \citep[FRB;][]{lorimer2007}, have been established as remarkable probes of physical processes spanning a vast range of scales: from atomic to cosmological \citep[e.g.][]{petroff2022}. While traversing plasma in different cosmic environments, the FRB signal experiences dispersion caused by the free electrons along the propagation path, resulting in a frequency-dependent time delay of the arrival of photons at telescopes on Earth. This characteristic 
dispersion, referred to as the dispersion measure (DM), has been used extensively to solve the so-called missing baryon problem \citep{mcquinn2014,macquart2020} and to infer how these baryons are distributed in the intergalactic (IGM) and circumgalactic (CGM) media \citep{simha2020, khrykin2024b, connor2025}. In addition, analysis of the FRB DM provides a tool for measuring cosmological parameters \citep[e.g.,][]{james2022b, wang2025, glowacki2024,fortunato2025} as well as for assessing galaxy feedback mechanisms \citep[e.g.][]{khrykin2024a, medlock2025, zzhang2025, guo2025, dong2025, kritti2025}.

\ctable[
caption = {Sample of \nfrb\ FRBs analysed in this work.},
star,
label = {tab:mtab},
width = 1.0\textwidth,
pos = t, center]{lccccrrc}
{
\tnote[]{\textbf{Notes.} From left to right, columns show full FRB name in TNS standard, right ascension, declination, FRB redshift, observed dispersion and rotation measures, total contribution to the rotation measure from the Milky Way (halo and ISM), and instrument or survey used to collect narrow-field spectroscopic information of  foreground galaxies.}
\tnote[a]{part of the FLIMFLAM DR1 \citep{khrykin2024b, huang2025}}
\tnote[b]{DM measurements adopted from \citet{bannister2019, macquart2020, heintz2020, bhandari2020} }
\tnote[c]{RM measurements adopted from \citet{scott2025}}
\tnote[d]{Milky Way RM contributions adopted from \citet{hutschenreuter2022}}
} 
{                   \FL
 FRB ID & R.A. & Dec. & Redshift & ${\rm DM_{obs}}$\tmark[b] & ${\rm \rmobs}$\tmark[c] & ${\rm RM_{MW}}$\tmark[d] & Source of Narrow-field Data \NN
        & ${\rm deg}$ & ${\rm deg}$ & & ${\rm pc~cm^{-3}}$ & ${\rm rad~m^{-2}}$  & ${\rm rad~m^{-2}}$ & \ML
 FRB~20211127I\tmark[a] & $199.8087$ & $-18.8380$ & $0.0469$ & $234.8$ & $-67.0\pm 1.0$ & $-2.9\pm6.2$ & 2dF-AAOmega, 6dF     \NN
 FRB~20211212A\tmark[a] & $157.3507$ & $+01.3605$ & $0.0707$ & $206.0$ & $+21.0\pm 7.0$ & $+6.0 \pm 5.7$ & 2dF-AAOmega \NN
 FRB~20190608B\tmark[a] & $334.0199$ & $-07.8983$ & $0.1178$ & $338.7$ & $+353.0\pm 1.0$ & $-24.4 \pm 13.3$ & SDSS, KCWI, MUSE\NN
 FRB~20200430A\tmark[a] & $229.7064$ & $+12.3768$ & $0.1610$ & $380.0$ & $-195.3\pm 0.7$ & $+14.5 \pm 7.$0 & LRIS, DEIMOS, MUSE \NN
 FRB~20210117A & $339.9795$ & $-16.1515$ & $0.2145$ & $729.2$ & $-45.8\pm 0.7$ & $+3.3 \pm 9.2$ & MUSE \NN
 FRB~20191001A\tmark[a] & $323.3513$ & $-54.7477$ & $0.2340$ & $506.9$ & $+51.1\pm 0.4$ & $+23.5 \pm 4.3$ & 2dF-AAOmega, MUSE \NN
 FRB~20210320C & $204.4608$ & $-16.1227$ & $0.2797$ & $384.6$ & $+288.8\pm 0.2$ & $-2.8 \pm 5.7$ & MUSE \NN
 FRB~20190102C & $322.4157$ & $-79.4757$ & $0.2912$ & $363.6$ & $-106.1 \pm 0.9$& $+26.6 \pm 7.7$ & MUSE \NN
 FRB~20180924B\tmark[a] & $326.1052$ & $-40.9000$ & $0.3212$ & $362.2$ & $+17.3\pm 0.8$ & $+16.5 \pm 5.0$ & 2dF-AAOmega, MUSE \NN
 FRB~20211203C & $204.5624$ & $-31.3803$ & $0.3439$ & $635.0$ & $+34.3\pm 1.2$ & $-29.2 \pm 9.1$ & MUSE \NN
 FRB~20200906A\tmark[a] & $53.4962$  & $-14.0832$ & $0.3688$ & $577.8$ & $+75.4 \pm 0.1$ & $+30.3 \pm 19.8$ & LRIS, DEIMOS, MUSE\NN
 FRB~20190611B & $320.7456$ & $-79.3976$ & $0.3778$ & $321.4$ & $+17.0\pm 3.0$ & $+29.0  \pm 10.8$ & MUSE \NN
 FRB~20181112A & $327.3485$ & $-52.9709$ & $0.4755$ & $589.3$ & $+10.5\pm 0.4$ & $+16.2 \pm 5.9$ & MUSE \NN
 FRB~20190711A & $329.4192$ & $-80.3580$ & $0.5220$ & $591.6$ & $+4.0\pm 1.0$ & $+19.4 \pm 6.5$ & MUSE 
\LL
}

Furthermore, if the propagation medium is magnetised, the intrinsic FRB polarization angle experiences a frequency-dependent rotation, which is characterised by the quantity dubbed the Faraday rotation measure (RM). Akin to the DM, the RM is proportional to the integrated number density of electrons along the propagation path, but additionally weighted by the magnetic field component parallel to the FRB sight line. It therefore offers an opportunity to probe the properties of the magnetic field on different scales in the Universe \citep{akahori2016, kovacs2024}. For instance, observed FRB RMs can be used to constrain the magnetic field generated by the FRB progenitors \citep[e.g.][]{piro2018,lyutikov2022,plavin2022,anna-thomas2023S}, to refine Milky Way magnetic field constraints  \citep{padhi2025}, or to shed light on the origin and strength of the magnetic field in the IGM, filaments, and voids of the cosmic web \citep[e.g.][]{hackstein2019,padmanabhan2023,mtchedlidze2024}.

 Measuring the strength of magnetic fields in galaxies is of particular interest, as these are interconnected with the processes of galaxy evolution and feedback \citep[e.g.][]{bertone2006,donnert2009,rodrigues2019}. FRBs provide a promising new approach to investigate galactic magnetism, its amplification, and its evolution. For instance, \citet{x2019} discovered that the FRB~$20181112$A sight line passes through the halo of a foreground galaxy at $b_{\rm impact}\simeq 29$~kpc. Through the analysis of the observed RM, they reported an upper limit on the corresponding parallel component of the magnetic field in the galactic halo of $B_{\parallel} \leq 0.8~\mu G$ (assuming fiducial halo parameters). This estimate is at odds with previous high-redshift measurements that analysed the RM from a sample of \ion{Mg}{II} absorbers at $\langle z \rangle = 1.3$ in quasar spectra and found $B_{\parallel} \approx 10~\mu G$ in the CGM \citep{bernet2008}. The small $B_{\parallel}$ value found by \citet{x2019} can indicate a little-magnetised halo or a highly disordered magnetic field. 

Recently, \citet{mannings2023} analysed a sample of nine FRBs and reported a positive correlation between DM and extragalactic \rmeg\ given by 

\begin{equation}
\rmeg = \rmobs - \rmmw~,
\label{eqn:RMeg}
\end{equation}
with \rmobs\ being the observed RM and \rmmw\ an estimate for the Milky Way contribution. They concluded that the majority of \rmeg\ thus must arise from the host galaxy. \citet{sherman2023} expanded the sample to 25 FRBs and came to a similar conclusion, emphasising that the bulk of extragalactic \rmeg\ is coming from the ISM of the FRB hosts. However, these works assumed that only FRB progenitors and/or the ISM contribute to the observer \rmobs, ignoring any contribution from the halos of the FRB host galaxies. Moreover, they did not account for potential contribution to the observed \rmobs\ from intervening foreground halos. While searching for foreground halos requires extensive spectroscopic observations, ignoring this information can lead to erroneous conclusions \citep[see][for a similar discussion about \dmfrb{} in FRB\,20190520B]{kglee2023}. 

The goal of this work is to constrain and disentangle the magnetic fields in all galactic environments traversed by the FRB sight lines, including intervening foreground galactic halos. In order to do this, we built a large spectroscopic dataset of galactic halos in the foreground of well-localised FRBs acquired by the FLIMFLAM collaboration \citep{kglee2022, simha2023, khrykin2024b, huang2025} and dedicated VLT/MUSE observations (PI: N. Tejos). We introduce a novel Bayesian Markov chain Monte Carlo (MCMC) statistical framework that takes into account both observational and modelling uncertainties and recovers the magnetic field strength in various media along the FRB sight lines with high precision levels.

This paper is organised as follows. In Section~\ref{sec:data}, we outline the main properties of the FRB dataset analysed in this work. We discuss our model for the observed RM and each contributing component in Section~\ref{sec:rm_model}. We summarise our Bayesian statistical model and present the results of the parameter inference in Section~\ref{sec:infernece}. We discuss our findings in Section~\ref{sec:disc} and conclude in Section~\ref{sec:end}. Throughout this work, we assumed a flat $\Lambda$CDM cosmology with dimensionless Hubble constant $h=0.673$, $\Omega_m = 0.315$, $\Omega_b=0.046$, $\sigma_8=0.8$, and $n_s=0.96$, consistent with the latest Planck results \citep{planck2018}.

\begin{figure*}
    \centering
    \includegraphics[width=0.8\textwidth]{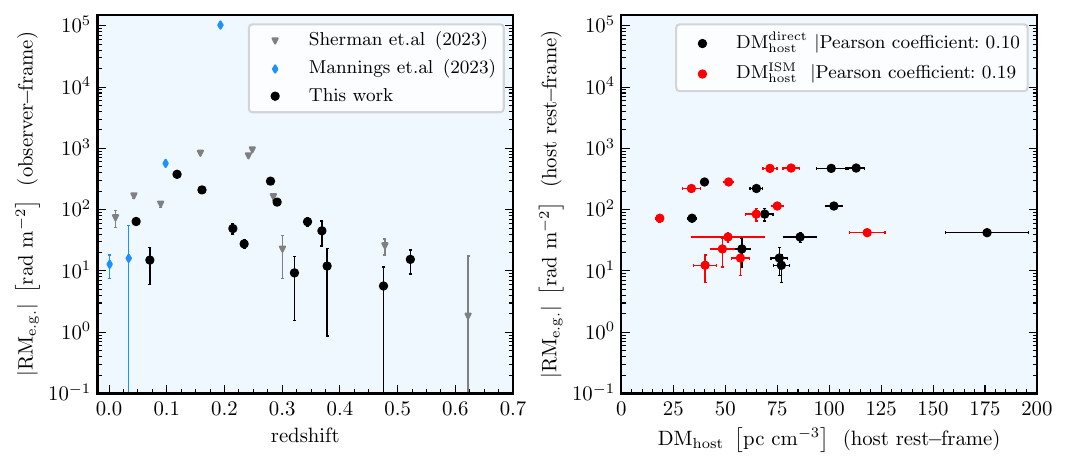}
    \caption{({\it Left}): Distribution of absolute values of \rmeg~as a function of FRB redshift in observer frame. The black dots show the FRB sample analysed in this work, whereas the grey triangles are the \texttt{DSA}-110 subsample taken from \citet{sherman2023} and the blue diamonds illustrate the FRBs from the \citet{mannings2023} sample, which are not part of this work. ({\it Right}): Distribution of \rmeg~values in rest-frame of FRB hosts as a function of the corresponding \dmhostdir\ and \dmhostism\ values from \citet{lucas2025} (only for sources for which corresponding MUSE data are available).}
    \label{fig:rm_vs_z}
\end{figure*}

\section{Data sample}
\label{sec:data}

In this work, we analysed a sample of \nfrb\ FRBs that were detected by the Commensal Real-time ASKAP Fast Transients \citep[\texttt{CRAFT};][]{macquart2010} survey conducted on the Australian Square Kilometre Array Pathfinder (ASKAP) radio telescope. The optical follow-up observations for host identification and redshift measurement were carried out as part of a collaboration between the \texttt{CRAFT} and Fast and Fortunate for FRB Follow-up\footnote{\url{https://sites.google.com/ucolick.org/f-4}} (${\rm F}^4$) teams. We built the sample according to the following set of criteria: 1) the posterior probability of FRB-host association by the Probabilistic Association of Transients to their Hosts (\texttt{PATH}; \citealt{aggarwal2021}) algorithm must be $P\left( O|x \right) \geq 0.90$; 2) the FRB must have a publicly available RM measurement; and 3) a given FRB field must have available spectroscopic observations of foreground galaxies. Only 14 FRBs to date satisfy all of the above criteria. We summarise the main properties of the FRBs in the sample in Table~\ref{tab:mtab}, noting that the \rmobs\ values adopted here are all taken from \cite{scott2025}.

The left hand panel of Figure~\ref{fig:rm_vs_z} shows the estimated extragalactic \rmeg, using Eq.~(\ref{eqn:RMeg})
and an estimate for \rmmw\ (see Table~\ref{tab:mtab}), in our sample as a function of the FRB redshift. Meanwhile, the right hand panel illustrates the distribution of \rmeg\ 
as a function of the estimated dispersion measure \dmhostdir\ associated with the host galaxies (halo+ISM/FRB progenitor) of the FRBs in the sample. We also show \dmhostism\ estimates separately via the red markers. The values are adopted from the analysis of H$\alpha$ flux and halo masses in the FRB hosts by \citet{lucas2025}. We ran a Pearson-correlation test between the \rmeg\ and \dmhostdir\ and between \rmeg\ and \dmhostism\ values and find no significant correlation in either case with $r_{\rm pearson}=0.10$ and $r_{\rm pearson}=0.19$, respectively \citep[see also][]{glowacki2025}.

We require foreground spectroscopic information in each FRB field in order to identify and characterise galaxies that are intersected by the FRB sight lines in our sample, and that might contribute to the observed \rmobs. Recently, the FLIMFLAM survey \citep{khrykin2024b,huang2025} acquired large spectroscopic samples of galaxies in the foreground of $8$ localized FRBs utilised in the FRB foreground-mapping approach \citep{kglee2022} to infer the distribution of cosmic baryons. For seven of the \nfrb\ FRBs in our sample, which are also part of FLIMFLAM DR1 \citep{khrykin2024b}, we adopted their narrow-field spectroscopic data \citep[within $10\arcsec$ of the FRB sight lines; see][]{huang2025}.  These observations were performed with the AAOmega spectrograph on the Anglo-Australian Telescope \citep{smith2004}, the Low-Resolution Imaging Spectrograph \citep[LRIS;][]{rockosi2010}, the DEep Imaging Multi-Object Spectrograph \citep[DEIMOS;][]{faber2003}, and the integral field unit (IFU) Keck Cosmic Web Imager \citep[\texttt{KCWI};][]{morrissey2018} on the Keck Telescopes at the W. M. Keck Observatory, and the IFU Multi-Unit Spectroscopic Explorer \citep[MUSE;][]{bacon2010} on the Very Large Telescope (VLT). We refer the interested reader to \citet{huang2025} for more detailed descriptions of these observations. For the remaining $7$ FRB fields in our sample lacking FLIMFLAM observations, we used \texttt{MUSE} to obtain the spectra of galaxies within the $1\times 1~{\rm arcmin}^2$ field of the respective FRB positions. Each FRB field has been observed for $4-8\times 800$~s with \texttt{MUSE} to reach $r \lesssim 25$ sources \citep{lucas2025}. 

\begin{figure*}
    \centering
    \includegraphics[width=0.9\textwidth]{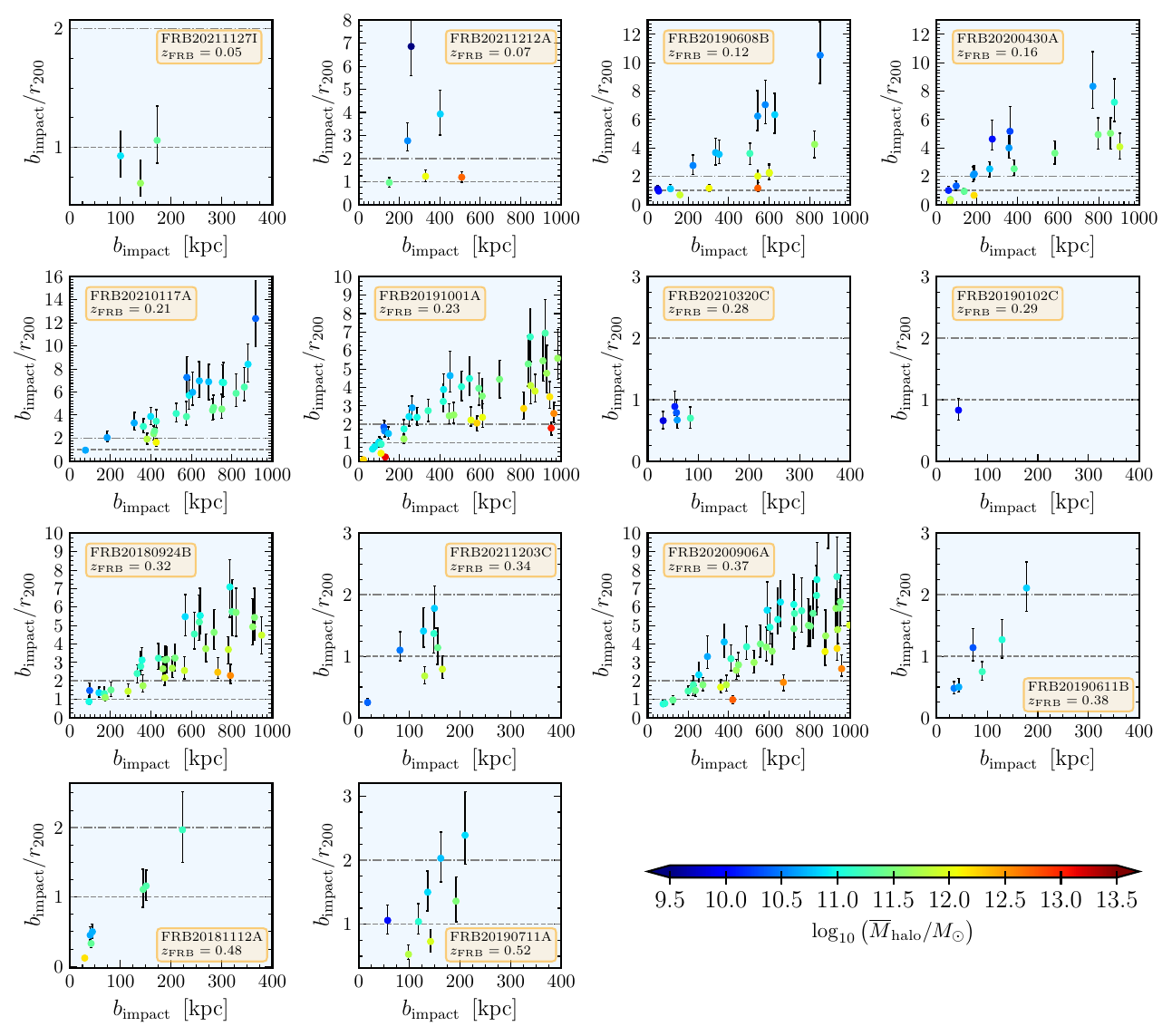}
    \caption{Impact parameter $b_{\rm impact}$ of identified foreground halos with respect to their virial radius, $r_{200}$, along each FRB sight line, plotted as a function of the impact parameter. The colour map illustrates the mean mass of the halos, whereas the associated uncertainty is propagated into the error bars on their corresponding $r_{200}$. The dashed line illustrates the maximum distance, $b_{\rm impact} = r_{200}$, at which a foreground halo can be intersected by the FRB sight line and potentially contribute to \rmobs\ if a halo extends to 1 virial radius, $r_{200}$. In addition, we also show the case where the halos extend to two virial radii, $2 \times r_{200}$, with the dash-dotted lines.}
    \label{fig:fg_distr}
\end{figure*}

Furthermore, we obtain photometric information on identified foreground galaxies by querying publicly available imaging surveys including the Dark Energy Survey \citep[\texttt{DES};][]{abbott2021}, the DECam Local Volume Exploration survey \citep[\texttt{DELVE};][]{dw2022}, and the the Panoramic Survey Telescope and Rapid Response System \citep[\texttt{Pan-STARRS};][]{chambers2016}. For \texttt{MUSE} sources without data in existing public imaging surveys, we constructed synthetic photometry adopting \texttt{SDSS} $g$, $r$, and $i$ filters, but manually setting the transmission to zero beyond the MUSE wavelength range of $4800-9300{\rm \AA}$. Additionally, we set the transmission to zero in the $5800-5960{\rm \AA}$ wavelength window in order to account for the blocking filter used to avoid the light from the laser guide stars. This photometric information is then used to estimate the corresponding stellar masses of identified galaxies via spectral-energy-distribution (SED) fitting. For this purpose, we adopted the publicly available SED-fitting algorithm \texttt{CIGALE}~\citep{boquien2019} with the initialisation parameters previously used by \citet{simha2023} and \citet{khrykin2024b}. In what follows, we adopt the mean stellar masses of foreground galaxies estimated by \texttt{CIGALE} and use the average stellar-to-halo mass relation of \citet{moster2013} to convert to the corresponding halo masses, $M_{\rm halo}$, at a given redshift.

Figure~\ref{fig:fg_distr} illustrates all halos of galaxies ($10 \leq \log_{10} \left( \overline{M_{\rm halo}}/M_{\odot}\right) < 12$) and groups ($12 \leq \log_{10} \left( \overline{M_{\rm halo}}/M_{\odot}\right) < 14$) that were observed in the foreground of each FRB in our sample. Those halos that are located at impact parameters $b_{\rm impact}$ less than equal to their respective virial radii $r_{200}$ can contribute to the observed \rmobs.
 
\section{Rotation measure model}
\label{sec:rm_model}

Rotation measure is one of the key observables of the FRB signal. It represents the change in the polarisation angle of the Faraday-rotated polarised emission propagated through the intervening ionised and magnetised gas. While the DM quantifies the integrated number density of free electrons along the propagation path of a signal, the RM describes the same integral, but weighted by the line-of-sight (parallel) component of the magnetic field, $B_{\parallel}$:

\begin{equation}
\label{eq:rmobs}
    \rmobs = \int \frac{B_{\parallel} n_e}{\left( 1 + z\right)^2} dz ~ .
\end{equation}

As an integral quantity,  \rmobs~can be described as a sum of contributions from several components along the FRB line of sight. Thus, for each $i$th FRB in our sample, we considered the following model for \rmobs~in 
the observer frame:

\begin{equation}
\label{eq:rmmodel}
    \rmmodel_{,i} = \rmmw_{,i} + \rmigm_{,i} + \sum_{j}^{\rm N_{fg}} \rmhaloij + \rmhost_{,i}~,
\end{equation}
where $\rmmw_{,i}$ corresponds to the contribution from the halo and ISM of the Milky Way (see Table~\ref{tab:mtab}); $\rmigm_{,i}$ arises from the diffuse gas in the IGM; $\rmhaloij$ is the contribution from the individual, intervening, foreground galactic halos; $N_{\rm fg}$ is the number of intersected foreground halos; and $\rmhost_{,i}$ comes from the FRB host galaxies. In what follows, we discuss each of these components in detail.

\begin{figure*}
    \centering
    \includegraphics[width=0.90\textwidth]{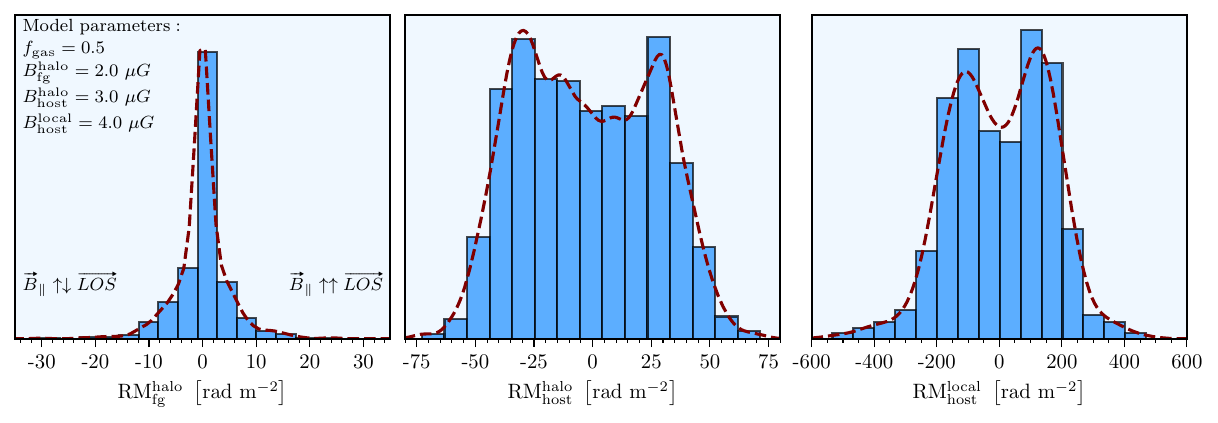}
    \caption{Example of typical distributions of RM values for FRB~$20211212{\rm A}$: ({\it left}) from the mNFW model of a single foreground galactic halo (see Section~\ref{sec:rm_halos});  ({\it middle}) from the mNFW model of the corresponding host-galaxy halo (see Section~\ref{sec:rm_host}); and ({\it right}) from the local environment in the ISM of the host galaxy and/or FRB progenitor (see Section~\ref{sec:rm_local}). The positive values of the histograms correspond to the RM values for the case when the line-of-sight component of the magnetic field is aligned with the direction of the FRB sight line (parallel), whereas the negative parts of the histograms, illustrate the case where the line-of-sight component of the magnetic field is directed away from the observer (anti-parallel). The dashed maroon lines show the KDE fit to the resulting RM distributions (see Section~\ref{sec:infernece}). Note the different range of RM values in the panels.}
    \label{fig:rm_parts}
\end{figure*}

\subsection{The Milky Way}
\label{sec:rm_mw}

The contribution from the Milky Way, \rmmw, arises both from the magnetised and ionised gas in the Galactic halo and from the ISM within the disk. In this work, we adopted the all-sky Faraday rotation map model from \citet{hutschenreuter2022} based on $\sim55\,000$ Faraday rotation sources compiled from multiple radio surveys (e.g. LOFAR Two-meter Sky Survey \citep{shimwell2017} and NRAO VLA Sky Survey \citep{Condon1998} RM catalogue). The map provides a \rmmw\ measurement with associated uncertainty for a given position on the sky. We list the resulting \rmmw\ values for each FRB field in Table~\ref{tab:mtab} \citep[see also][]{padhi2025}.

\subsection{The intergalactic medium}
\label{sec:rm_igm}

Due to expected multiple reversals of the intergalactic magnetic field, the \rmigm, arising from the low-density IGM gas tracing the large-scale structure (LSS) of the Universe, is expected to be small relative to other contributing environments (see Eq.~\ref{eq:rmmodel}). The recent LOFAR measurement of $\langle {\rm RM} \rangle \simeq 0.71 \pm 0.07~{\rm rad~m^{-2}}$ in LSS filaments is consistent with  magnetic field of $\simeq 32$~nG \citep[e.g.][and the references within]{amaral2021,pomakov2022,carretti2022,carretti2023}. Given that this \rmigm~value is at least $1-2$ orders of magnitude smaller than the \rmobs~and \rmmw~(see Table~\ref{tab:mtab}), for the remainder of the paper we assume that the IGM contribution is negligible; i.e. our model adopts

\begin{equation}
    \rmigm = 0 \, \rmunits.
\end{equation}  

\subsection{The foreground halos}
\label{sec:rm_halos}

In order to estimate the electron density and RM contribution of a given foreground halo intersected by the FRB sight line, we utilised the modified Navarro-Frank-White model \citep[mNFW;][]{xyz2019}. This model yields a radial density profile $\rho_b\left(r\right)$ of the gas in individual halos given by

\begin{equation}
\label{eq:rhob}
    \rho_b\left( r\right) = \fgas \frac{\Omega_b}{\Omega_m}\frac{\rho_0\left( M_{\rm halo}\right)}{y^{1-\alpha} \left( y_0 + y\right)^{2+\alpha}}~,
\end{equation}
where \fgas\ is the fraction of cosmic baryons located in the ionised CGM relative to the total amount of baryons within individual galactic halos; $\rho_0$ is the central density of the halo as a function of halo mass, $M_{\rm halo}$; $y\equiv c\left( r/r_{200} \right)$, where $c$ is the concentration parameter; while $y_0$ and $\alpha$ are mNFW model parameters. Here, we adopted the fiducial values $y_0=\alpha=2$ from \citet{xyz2019} while considering \fgas\ as a free parameter. All identified foreground galaxies and groups are illustrated in Figure~\ref{fig:fg_distr}, where we plot the ratio of the halo impact parameter to the corresponding virial radius, $r_{200}$, as a function of the impact parameter. 

The ${\rm RM_{fg}^{halo}}$ contribution of an individual halo in units of \rmunits\ is obtained by integrating the density profile $\rho_{b}\left(r\right)$ given by eq.~(\ref{eq:rhob}) as 

\begin{equation}
\label{eq:rm_halo}
{\rm RM_{fg}^{halo}} = \frac{\bhalos~\alphafh~e^3\mu_e}{2\pi m_e^2c^4 m_{p}\mu_{H}} 2\int_{0}^{\sqrt{r_{\rm max}^2 - r_{\perp}^2}} \rho_b\left( s \right) {\rm d}s ~,
\end{equation}
\cite[e.g.,][]{akahori2011}, where $e$ and $m_e$ are the electric charge and mass of the electron, $c$ is the speed of light, $m_{p}$ is the mass of the proton, $\mu_H = 1.3$ and $\mu_e=1.167$ take into account both hydrogen and helium atoms; $r_{\perp}$ is the impact parameter of the FRB sight line with respect to the centre of the halo, $r_{\rm max}$ is the maximum extent of the halo in units of $r_{200}$, and $s$ is the path that the FRB follows inside a given halo. In what follows, we adopted a fixed $r_{\rm max} \equiv r_{200}$ value, similarly to the approach of \citet{khrykin2024b}\footnote{In principle, $r_{\rm max}$ should also be included as another free parameter \citep[see][]{xyz2019,simha2020,kglee2023}. However, we tested the model using $r_{\rm max} =2\times r_{200}$ and did not find significant differences in the results. Therefore, in this work, we adopted a fixed truncation radius and will explore a more sophisticated parametrisation in the future.}. The term \bhalos~describes the average strength of the total magnetic field in the foreground galactic halos in units of $\mu G$, and it is one of the free parameters we considered in this work. In what follows, we assume \bhalos\ is constant with respect to $r$. Furthermore, \bhalos~is multiplied by the cosine of the angle $\alpha_{\rm fg}$ between the direction of the magnetic field vector and line of sight in order to obtain the required parallel component (see Eq.~\ref{eq:rmmodel}). We note that ${\rm RM_{fg}^{halo}}$ in eq.~(\ref{eq:rm_halo}) is defined in the rest-frame of each foreground halo. We therefore additionally multiplied Eq.~(\ref{eq:rm_halo}) by a $\left( 1 + z_{\rm halo}\right)^{-2}$ to account for the redshift dilation and convert the result to the observer frame.

In addition, both the stellar mass estimate and the stellar-to-halo mass relation are subjects of considerable uncertainties. Similarly to \citet{khrykin2024b}, we took these uncertainties into account by introducing a random scatter of $0.3$~dex to the inferred mass of each halo (see \citealp{simha2021} for more details). For each foreground halo, we constructed a log-normal distribution of its halo mass adopting the corresponding mean and standard deviation values. We then randomly drew $N=1000$ realisations of the halo mass from this distribution. Additionally, in order to account for the orientation of the magnetic field, for each draw we randomly chose an angle, $\alpha_{\rm fg}$, from the uniform distribution $\mathcal{U}\left[ 0, \pi \right]$ and calculated the corresponding \rmhalo~value given by Eq.~(\ref{eq:rm_halo}). 

An example of the model distribution of \rmhalo~values for a single foreground halo located $b_{\rm impact}=150$~kpc away from the FRB~$20211212$A sight line is illustrated by the histogram in the left panel of Figure~\ref{fig:rm_parts}. It is apparent that the \rmhalo~histogram is centred around ${\rm RM}\simeq 0~{\rm rad~m^{-2}}$. In the particular case of this foreground halo, for the majority of the halo mass realisations the extent of the halo, $r_{200}$, is smaller than the corresponding impact parameter of the halo ($b_{\rm impact}=150$~kpc; see Figure~\ref{fig:fg_distr}). Therefore, in the realisations where the sight line does not actually intersect the foreground halo, the resulting $\rmhalo$~is effectively zero.

\subsection{The FRB host galaxy}

The contribution from the FRB host galaxies to the observed \rmobs~remains highly uncertain. Previous studies analysing the FRB RMs did not have information about the foreground large-scale structures (LSS) and galactic halos, attributing all the extragalactic RM (after subtracting the contribution from the Milky Way) to the FRB hosts. Moreover, some previous works assumed that all of the host contribution arises only from the ISM or local FRB environment, ignoring potential contributions from the extended halo of the host galaxy \citep[e.g.][]{mannings2023}. In this work, we attempted to explicitly model both the halo and local contributions from the FRB hosts, employing the following model:

\begin{equation}
    \rmhost = \frac{\rmhhost + \rmlhost}{(1 + z_{\rm host})^2}~,
\end{equation}
where \rmhhost~describes the contribution from the halo of the FRB host galaxy, and \rmlhost~denotes the contribution from the FRB progenitor itself and/or the ISM environment of the host galaxy, in which the progenitor resides. Here, similarly to the foreground galaxies, we rescaled \rmhost~to the observer frame by a factor of $\left( 1+z_{\rm host}\right)^{-2}$. In what follows, we describe the calculation of each of these terms separately. 

\subsubsection{The host halo}
\label{sec:rm_host}

In order to estimate the \rmhhost~for a given FRB in our sample, we followed the discussion in Section~\ref{sec:rm_halos} and similarly adopted the mNFW model. We utilised publicly available estimates of the host stellar masses from \cite{gordon2023} to describe their halo radial density profiles $\rho_b\left( r\right)$. We then obtained the value of the \rmhhost~by integrating $\rho_b\left( r\right)$ using Eq.~\ref{eq:rm_halo}. However, contrary to the \rmhalo~case, we assumed that FRB is located inside the disc of the host galaxy and only integrated over the half of the host halo as 

\begin{equation}
\label{eq:rm_hhost}
{\rm RM_{host}^{halo}} = \frac{\bhosts~\alphahh~e^3\mu_e }{2\pi m_e^2c^4m_{p}\mu_{H}} \int_{0}^{\sqrt{r_{\rm max}^2 - r_{\perp}^2}} \rho_b\left( s \right) {\rm d}s~,
\end{equation}
where \bhosts~is the the average strength of the total magnetic field in the halos of the FRB hosts in units of $\mu G$, and it is another free parameter considered in this work. Similar to discussion in Section~\ref{sec:rm_halos}, we assume \bhosts\ is constant with respect to location $r$ in the halo. $\alpha_{\rm hh}$ is the angle between the direction of the \bhosts~vector and the line of sight. 

We took into account the uncertainty in the halo masses by introducing a random $0.3$~dex scatter and performed a Monte Carlo sampling ($N=1000$~draws) of the corresponding log-normal distribution of the FRB hosts' halo masses. Similarly to the \rmhalo~calculations in Section~\ref{sec:rm_halos}, for each halo mass realisation we drew a value of $\alpha_{\rm hh}$ from the corresponding uniform distribution $\alpha_{\rm hh} \in \mathcal{U}\left[ 0, \pi \right]$ to take into account the orientation of the magnetic field relative to the line of sight. This procedure results in a distribution of \rmhhost~values for each host in the sample. An example of such a distribution for the halo of the FRB~$20211212$A host galaxy is illustrated in the middle panel of Figure~\ref{fig:rm_parts}. 

\subsubsection{The host ISM/FRB progenitor}
\label{sec:rm_local}

To describe the RM associated either with the FRB source itself or with the host galaxy ISM (or both), which we term \rmlocal, we utilised the RM-DM relation \citep{akahori2016, pandhi2022}, which after accounting for all the constants is given by (in host rest-frame)

\begin{equation}
\label{eq:rm_local}
    \rmlocal = \frac{\blocal~\alphahl~\dmlocal }{1.22},
\end{equation}
where \dmlocal~is the rest-frame dispersion measure in the local FRB environment in units of ${\rm pc~cm^{-3}}$, the \blocal~is the average strength of the total magnetic field associated with ISM/progenitor in units of $\mu G$, and $\alpha_{\rm hl}$ is the angle between the direction of the \blocal~vector and the line of sight. In what follows, we consider \blocal~as another free parameter in our model.

Recently, the analysis of the FLIMFLAM DR1 \citep{khrykin2024b, huang2025} yielded the first ever published direct estimate of \dmlocal~(DM contribution from the FRB progenitor and/or host ISM), inferring an average contribution of $\dmlocal = 69^{+28}_{-19}~{\rm pc~cm^{-3}}$. We adopted these findings and drew $N=1000$ realisations of \dmlocal~from the corresponding 1D posterior PDF from \citet{khrykin2024b}. Similarly to what is highlighted in the discussion in Sections~\ref{sec:rm_halos} and \ref{sec:rm_host}, we sampled the uniform distribution $\alpha_{\rm hl} \in \mathcal{U}\left[0, \pi\right]$ in order to take into account the orientation of the \blocal~vector relative to the FRB sight line. The resulting \rmlocal~distribution for FRB~$20211212$A, calculated using Eq.~(\ref{eq:rm_local}), is illustrated by the histogram in the right hand panel of Figure~\ref{fig:rm_parts}.

\section{Statistical inference}
\label{sec:infernece}

In order to estimate the set of model parameters $\Theta = \{ \fgas, \bhalos, \bhosts, \blocal \}$ and associated uncertainties, we adopted a Bayesian inference formalism. Therefore, first, we need to define the Bayesian likelihood function $\mathcal{L_{\mathrm{FRB}}}\left( \rmobs_{,i} | \Theta \right)$ for the observed $\rmobs_{,i}$ given $\Theta$ for each $i\mathrm{th}$ FRB in the sample.

\subsection{The likelihood function}
\label{sec:like}

We began by constructing a grid of parameter values on which the likelihood will be estimated. For \fgas~, we adopted a range of $\fgas = \left[ 0.01, 1.0\right]$ with the step $\Delta=0.1$. Further, we assumed the same range for the strength of the magnetic field in three environments $B \equiv \bhalos \equiv \bhosts \equiv \blocal = \left[ 0, 20 \right]~\mu {\rm G}$ with step $\Delta=0.5~\mu {\rm G}$.  

Having defined the parameter grid, we obtain the values of the likelihood function, $\mathcal{L_{\mathrm{FRB}}}\left( \rmobs_{,i} | \Theta \right)$, for a single $i$th FRB at each point of the parameter grid following the algorithm from \citet{khrykin2021} as follows.

\begin{figure}
    \centering
    \includegraphics[width=\columnwidth]{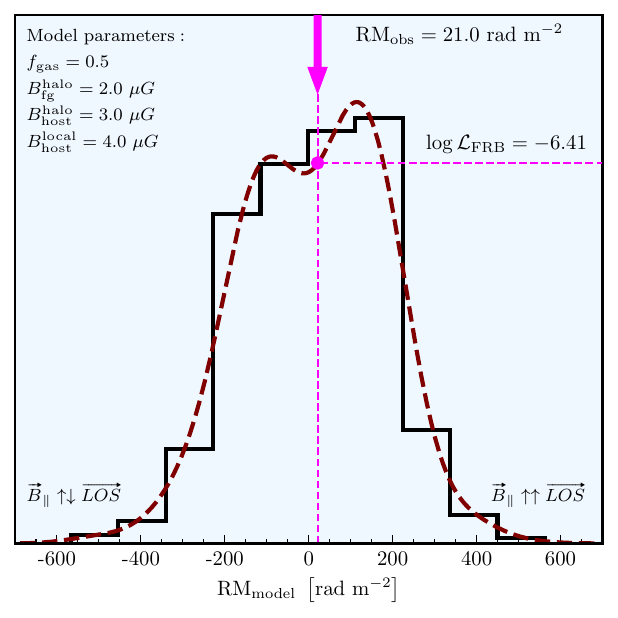}
    \caption{Resulting distribution of \rmmodel~values given by Eq.~(\ref{eq:rmmodel}), estimated by combining contributions from each environment along the FRB sight line (see Section~\ref{sec:rm_model} and Figure~\ref{fig:rm_parts} for details). The vertical dashed magenta line illustrates the value of the \rmobs~for FRB~$20211212$A, while the horizontal magenta dotted line shows the value of the corresponding log-likelihood value ($\log\mathcal{L_{\rm FRB}} = -6.41$) evaluated from the KDE fit to the \rmmodel~distribution at the value of \rmobs, given the combination of model parameters $\Theta$ (see discussion in  Section~\ref{sec:infernece} for more details).}
    \label{fig:rm_model}
\end{figure}

\begin{enumerate}
    \item Given the \bhalos\ and \bhosts\ values in $\Theta$, we calculated the corresponding distributions of $\rmhalo_{,i}$ and $\rmhhost_{,i}$, respectively (see Section~\ref{sec:rm_halos} and \ref{sec:rm_host}). Note that for each foreground or FRB host halo, we performed a new random draw from the corresponding uniform distribution describing the orientation angles of a given magnetic field vector. We fitted each of these distributions with a kernel density estimator (KDE; maroon dashed lines in the left hand and middle panels of Figure~\ref{fig:rm_parts}) and then resampled them by randomly drawing $N=2000$ realisations of \rmhalo\ and \rmhhost\ values.  

    \item Given the value of \blocal~and assuming the \texttt{FLIMFLAM} estimates of \dmlocal\, we calculated a distribution of $\rmlocal_{,i}$ values following Eq.~(\ref{eq:rm_local}). Similarly, to the previous step (1), we fitted the resulting distribution with the KDE (dashed maroon line in the right hand panel of Figure~\ref{fig:rm_parts}) and randomly drew $N=2000$ values of \rmlocal. 

    \item Given the value of $\rmmw_{,i}$ and the associated $1\sigma$ uncertainty (see Table~\ref{tab:mtab}), we constructed a corresponding Gaussian distribution from which we randomly drew $N=2000$ realisations of \rmmw.

    \item We calculated the joint distribution of $\rmmodel_{,i}$ values given by Eq.~(\ref{eq:rmmodel}) by adding together $N=2000$ ${\rm RM}$ samples obtained through steps (1)-(3). An example of such a joint $\rmmodel_{,i}$ distribution for a single realisation of $\Theta$ for FRB~$20211212$A is shown by the histogram in Figure~\ref{fig:rm_model}.

    \item We applied the KDE to the resulting joint $\rmmodel_{,i}$ distribution to find its continuous probability-density function (PDF; see dashed maroon line in Figure~\ref{fig:rm_model}). Finally, we estimated the value of the likelihood function by evaluating the resulting total KDE PDF at the observed value of the $\rmobs_{,i}$ for a given FRB (see Table~\ref{tab:mtab} and vertical magenta line in Figure~\ref{fig:rm_model}). 

    \item We repeated steps (1)-(5) for each combination of model parameters, $\Theta$.
\end{enumerate}

This procedure results in $N=N_{f_{\rm gas}} N_{B}^3 = 758131$ determinations of the likelihood function $\mathcal{L_{\mathrm{FRB}}}\left( \rmobs_{,i} | \Theta \right)$ evaluated on $\Theta \equiv \{ \fgas, \bhalos, \bhosts, \blocal \}$ parameter grid for each FRB in the sample. Next, we obtained a joint-likelihood function for the entire sample by taking the product of the individual (independent) likelihood functions: 

\begin{equation}
\label{eq:jlf}
\mathcal{L}_{\rm joint} = \prod_i^{N_{\mathrm FRB}} \mathcal{L_{\mathrm{FRB}}}\left( \rmobs_{,i} | \Theta \right),
\end{equation}
where $N_{\rm FRB}=\nfrb$ is the number of FRBs in our sample (see Table~\ref{tab:mtab}).

Finally, we interpolated $\mathcal{L}_{\rm joint}$ given by Eq.~(\ref{eq:jlf}) for any arbitrary combination of parameter values that lie between the grid points of the parameter space. 

\subsection{The priors}
\label{sec:priros}
 
As in any Bayesian inference, it is important to be explicit about the adopted priors $\pi$ on our model parameters, which we define hereafter. First, given the lack of observational information, we adopted a flat, uniform, non-informative prior on the fraction of cosmic baryons in the galactic halos; \fgas, i.e., $\pi\left( \fgas \right) = \left( 0, 1 \right]$. Similarly, we used a flat uniform prior for each of the magnetic fields, $\pi\left( \bhalos \right) = \pi \left( \bhosts\right) = \pi \left( \blocal \right) = \left[ 0, 20 \right]~{\rm \mu G}$ \citep[e.g.][]{beck2013}. Moreover, given the difference in the densities in the halos and the ISM environment of the galaxies, we set the following extra condition: $\bhalos < \blocal$ (and, analogously, $\bhosts < \blocal$). This is consistent with the expectations from magneto-hydrodynamic simulations \citep[e.g.][]{ramesh2023}.

\begin{figure}
    \centering
    \includegraphics[width=\columnwidth]{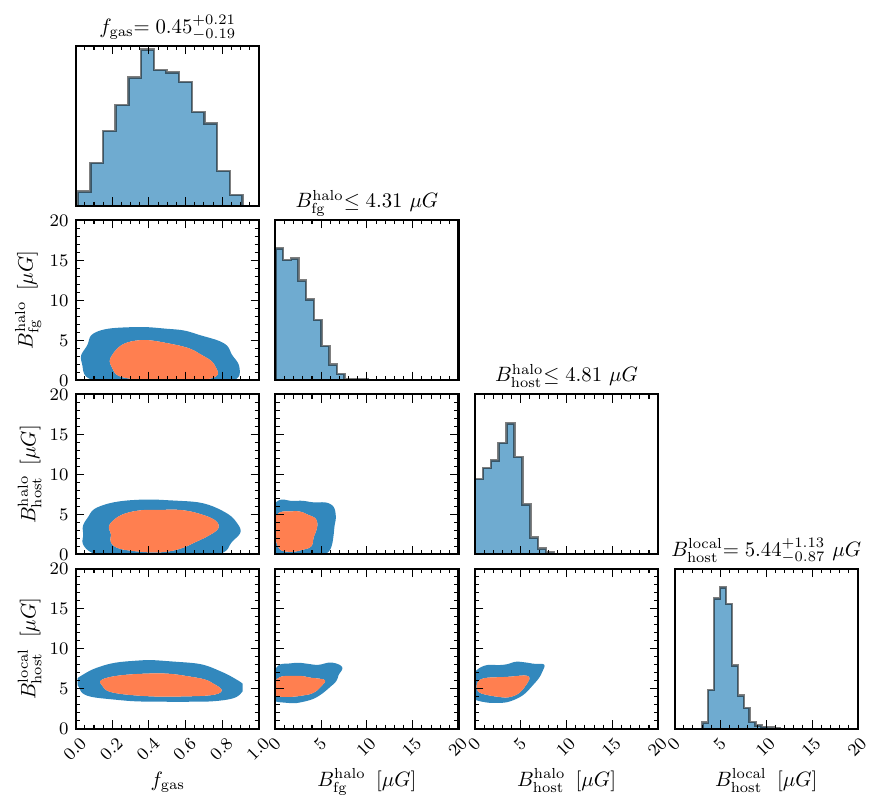}
    \caption{MCMC inference result for sample of \nfrb~FRBs in our dataset (see Table~\ref{tab:mtab}). The orange (blue) contours correspond to the inferred $68\%$ ($95\%$) confidence intervals. The diagonal panels show the corresponding marginalised $1$D posterior probabilities of the model parameters.}
    \label{fig:mcmc_data}
\end{figure}

In addition, following \cite{khrykin2024b} we adopted a prior to the total budget of cosmic baryons in the diffuse states outside of the individual galaxies, \fdiff, as 

\begin{equation}
\label{eq:fdz}
    \fdiff\left( z \right) \equiv \figm + \fcgm + \ficm = 1 - f_{\rm stars} - f_{\rm bh} - f_{\rm ism}~,
\end{equation}
where \figm\ is the fraction of cosmic baryons residing in the diffuse IGM gas, \ficm\ is the fraction of cosmic baryons inside the intracluster medium (ICM) of galaxy clusters with $M_{\rm halo} \geq 10^{14}~M_{\odot}$, while \fcgm\ is the cosmic baryon fraction in all the halos at $M_{\rm halo} < 10^{14}~M_{\odot}$; in contrast to \fgas, which represents the average fraction of the ionised CGM baryons in the individual halos, \fcgm\ represents the fraction of baryons integrated over all CGMs in the Universe. We followed the discussion of \citet{macquart2020} to estimate  $\fdiff\left( z \right)$, and we refer the interested reader to the full description provided by \citet{khrykin2024b}. Given the uncertainties in the stellar IMF, evaluating $\fdiff\left( z\right)$~at the mean redshift of our FRB sample, $\langle z_{\rm sample} \rangle \simeq 0.27$, yields $\fdiff\left( z=0.27\right) \simeq 0.86 \pm 0.02$. 

In the following, we assumed $\figm=0.59\pm 0.10$, which is the value found in the FLIMFLAM DR1 analysis by \cite{khrykin2024b}. Following the approach in that work,  we calculated the look-up conversion table between \fgas~and \fcgm~by integrating the \texttt{Aemulus} halo mass function \citep{mcclintock2019} over the mass range of the halos in our sample, $10 \lesssim \log_{10}\left( M_{\rm halo} / M_{\odot} \right) \lesssim 13.1$ (this includes both foreground and FRB host halos). A similar conversion is adopted between $f_{\rm gas}^{\rm icm}$ and \ficm~assuming $M_{\rm halo} \geq 10^{14}M_{\odot}$ and $f_{\rm gas}^{\rm icm} = 0.8\pm 0.1$, which is consistent with measurements of gas in galaxy clusters \citep{gonsalez2013,chiu2018}. 

\begin{figure*}
    \centering
    \includegraphics[width=\textwidth]{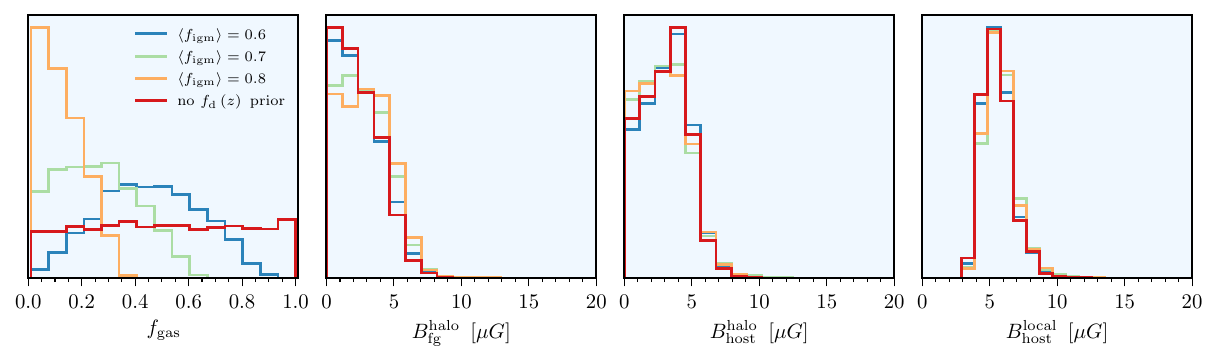}
    \caption{1D marginalised posterior probability distributions of model parameters considered in this work, estimated with different values of $\langle \figm \rangle$ adopted for the \fdiffz~prior (see Section~\ref{sec:priros}). The fiducial model with $\langle \figm \rangle = 0.60$ is shown by the blue curves.}
    \label{fig:ppdfs}
\end{figure*}

Additionally, we note that our sample of foreground (see Figure~\ref{fig:fg_distr}) and FRB hosts' halos does not cover the full range of halo masses below the cluster mass limit of $M_{\rm halo} \simeq 10^{14}M_{\odot}$. Therefore, we split \fcgm~into $f_{\rm cgm}^{\rm sample}$, representing the halo mass range of our sample, and $f_{\rm cgm}^{\rm other}$ corresponding to the halo mass range of $13.1 \leq \log_{10}\left( M_{\rm halo} / M_{\odot} \right) < 14.0$ not covered by our data. Consequently, at each step of the MCMC inference, the proposed value of \fgas~is converted to $f_{\rm cgm}^{\rm sample}$~using the precomputed lookup table. For the non-cluster halo mass range not covered by our data, we randomly picked a value of $f_{\rm gas}$ for these halos from the uniform distribution $\mathcal{U}\left(0,1 \right]$ and converted into the corresponding $f_{\rm cgm}^{\rm other}$ fraction also adopting the \texttt{Aemulus} halo mass function. These terms, together with the mean value of \ficm, are then compared to the range of \fdiffz~in Eq.~(\ref{eq:fdz}), yielding the following prior:

\begin{equation}
    \langle \fdiff \rangle -\sigma_{\rm tot} \leq \langle f_{\rm igm} \rangle + f_{\rm cgm}^{\rm sample} + f_{\rm cgm}^{\rm other} + \langle \ficm \rangle \leq \langle \fdiff \rangle + \sigma_{\rm tot}~,
\end{equation}
where $\sigma_{\rm tot}$ takes into account the corresponding uncertainties in \fdiff, \figm, and \ficm, and is given by

\begin{equation}
    \sigma_{\rm tot} = \sqrt{ \sigma_{\fdiff}^2 +  \sigma_{\figm}^2 + \sigma_{\ficm}^2 } \approx 0.14.
\end{equation}

\subsection{Inference results}
\label{sec:results}

Given the expression for the joint-likelihood function described in Section~\ref{sec:like} and the choice of priors in Section~\ref{sec:priros}, we proceeded to sample this $\mathcal{L}_{\rm joint}$. We adopted the publicly available affine-invariant MCMC sampling algorithm \texttt{EMCEE} by \cite{fm2013} to obtain the posterior probability distributions for our model parameters.

The resulting posterior PDFs of the four model parameters $\Theta \equiv \{ \fgas, \bhalos, \bhosts, \blocal \}$ are illustrated in Figure~\ref{fig:mcmc_data}, where the orange (blue) contours correspond to the $68\%$ ($95\%$) confidence intervals (CIs), whereas the marginalised 1D posterior PDFs for each of the model parameters are shown in the diagonal top panels. It is apparent upon inspection of the marginalised posterior PDF that, for some of the considered model parameters, we can only place upper limits, whereas for the others we are able to make a measurement. In order to distinguish between these two cases, we adopted the following criterion. If the peak of the posterior PDF is significantly larger than the posterior PDF values at the edges of the corresponding parameter range, then we classified it as a measurement. We then quoted the median $50{\rm th}$ percentile of the marginalised posterior distributions as the measured values, while the corresponding uncertainties are estimated from the $16{\rm th}$ and $84{\rm th}$ percentiles, respectively. On the other hand, if the above criterion was not met, we reported an upper limit by quoting the $84$th percentile (effectively $1\sigma$) of the posterior PDF for a given parameter.

Given our choice of priors in Section~\ref{sec:priros}, we estimated the average strength of the magnetic field inside the ISM and/or FRB progenitor to be $\blocal = 5.4^{+1.1}_{-0.9}~\mu {\rm G}$. On the other hand, we were only able to place an upper limit on the average strength of the magnetic field in the FRB host halos because the $\bhosts \simeq 0.0~\mu {\rm G}$ has a significant posterior probability. Our inference yields $\bhosts \lesssim 4.8~\mu G$. Similarly, we placed an upper limit on the average strength of the magnetic field in foreground halos, $\bhalos \lesssim 4.3~\mu G$. 
Finally, we inferred the fraction of cosmic baryons inside individual galactic halos to be $\fgas = 0.45^{+0.21}_{-0.19}$. Following the discussion in Section~\ref{sec:priros} \citep[see][for more details]{khrykin2024a}, this implies that a fraction $f_{\rm cgm}^{\rm sample} = 0.14^{+0.07}_{-0.06}$ of cosmic baryons reside inside the CGM of the $10 \lesssim \log_{10}\left( M_{\rm halo}/M_{\odot}\right) \lesssim 13.1$ halos.

We present a validation of the accuracy of our inference procedure based on mock FRB experiments in Appendix~\ref{sec:appendix}, to which we refer the interested reader.

\section{Discussion}
\label{sec:disc}

We now examine how the results of the inference depend on the assumed choice of priors, and discuss our findings in a broader context of previous extragalactic $B_{\parallel}$ measurements.

\subsection{Effect of the IGM baryon fraction}
\label{sec:disc_figm}

In our analysis, we explicitly assume that IGM contribution to the \rmobs~is negligible and do not consider \figm~as a free parameter in our model (see discussion in Section~\ref{sec:rm_igm}). Nevertheless, we needed to adopt a value of \figm\ as part of the prior for our inference algorithm. However, there is a certain degree of degeneracy between \figm~and \fgas, set by the \fdiffz~prior (see Eq.~\ref{eq:fdz}), with the uncertainty due to the fact that the exact partition of the cosmic baryons between the diffuse gas in the IGM and the ionised CGM remains uncertain. Recent FRB studies have produced a mild $1\sigma$ tension in the inferred \figm~fractions. \citet{khrykin2024b} presented the first results of the FLIMFLAM survey, which utilises a density reconstruction algorithm to constrain the LSS distribution in the foreground of $8$ localized FRBs. Their analysis yielded $\figm = 0.59\pm 0.10$, mildly disfavouring strong AGN feedback \citep{khrykin2024a}. On the other hand, \citet{connor2025} calibrated the observed DM distribution of $69$ localised FRBs (but without information on the foreground halos) to the \texttt{TNG}~300 hydrodynamical simulations. They found $\figm = 0.76^{+0.10}_{-0.11}$ (see also \citealp{hussaini2025}), pointing to stronger feedback. In general, this discrepancy in the inferred \figm~values might affect the accuracy of our inference (i.e. a systematic effect). In order to quantify the significance of the \figm~constraint on the inference results, we repeated our MCMC analysis described in Section~\ref{sec:infernece}, but with different values of the mean \figm~fraction adopted in the \fdiffz~prior (see Section~\ref{sec:priros}). The resulting $1$D marginalised posterior distributions of the model parameters $\Theta$ are illustrated in Figure~\ref{fig:ppdfs}, where we show inference results for adopted $\langle \figm \rangle = 0.6$ (blue), $\langle \figm \rangle = 0.8$ (orange), and the intermediate case $\langle \figm \rangle = 0.7$ (green). In addition, we also show the MCMC results without \fdiffz~prior (red curves).

It is apparent from Figure~\ref{fig:ppdfs}, that the choice of $\langle \figm \rangle$ value significantly affects the resulting constraints on \fgas. Indeed, for $\langle \figm \rangle = 0.8$, we estimate $\fgas \lesssim 0.22$ ($f_{\rm cgm}^{\rm sample} \lesssim 0.07$), which is at least $\simeq 2$ times lower than in our fiducial case with $\langle \figm \rangle = 0.6$; i.e. $\fgas = 0.45^{+0.21}_{-0.19}$. This behaviour is expected given the functional form of the \fdiffz~prior described in Eq.~(\ref{eq:fdz}), in which \figm{} and \fgas{} are approximately complements of each other.
Interestingly, the choice of $\langle \figm \rangle$ does not substantially change the constraints on the \bhalos\ and \bhosts, which are illustrated in the middle panels of Figure~\ref{fig:ppdfs}. Both of these quantities are linearly correlated with \fgas, given by Eq.~(\ref{eq:rm_halo}) and Eq.~(\ref{eq:rm_hhost}). 
Nevertheless, we observe that even large changes in $\langle \figm \rangle$ and the correspondingly inferred \fgas~imply inferences for \bhalos\ between $\bhalos \lesssim 4.3~\mu G$ ($\langle \figm \rangle = 0.6$) and $\bhalos \lesssim 4.9~\mu G$ ($\langle \figm \rangle = 0.8$), and that \bhosts\ and \blocal\ remain mostly unaffected. This is expected given that the magnetic field associated with the ISM of the host galaxy or FRB progenitor does not depend on the diffuse cosmic baryons (see Eq.~\ref{eq:rm_local}).

According to the results illustrated in Figure~\ref{fig:ppdfs}, improving constraints on \figm\ will be crucial for obtaining the precise measurement on \fgas, and therefore for constraining the galactic feedback models \citep{ayromlou2023, khrykin2024a, medlock2024, connor2025, leung2025}. The upcoming FLIMFLAM DR2 (Simha et al. in prep) will incorporate the analysis of $\approx 25$ localised FRBs with the mapping of foreground structures, which should be enough to bring the \figm\ uncertainty down to the $\approx 5\%$ level. Additionally, we note that throughout this work we adopted a single \fgas~parameter to describe the cosmic baryon fraction in all individual halos within a mass range of $10 \lesssim \log_{10}\left( M_{\rm halo} / M_{\odot}\right) \lesssim 13.1$, which our data constrain. However, hydrodynamic simulations indicate that \fgas\ is not only sensitive to the feedback prescriptions, but it is also a function of the halo mass \citep{khrykin2024a}. The FLIMFLAM-like inference would require $N_{\rm FRB}\simeq 300$ to distinguish between the \fgas\ fractions in different halo mass bins of the $\approx 10\%$ level \citep{huang2025}. The analysis presented in this work suggests that utilizing the RM information alone or in conjunction with FRB DMs and the LSS density reconstructions can dramatically reduce the number of FRBs required to make similar precision predictions on the \fgas~parameter as a function of halo mass. We will explore this in future work.

\begin{figure*}
    \centering
    \includegraphics[width=0.9\textwidth]{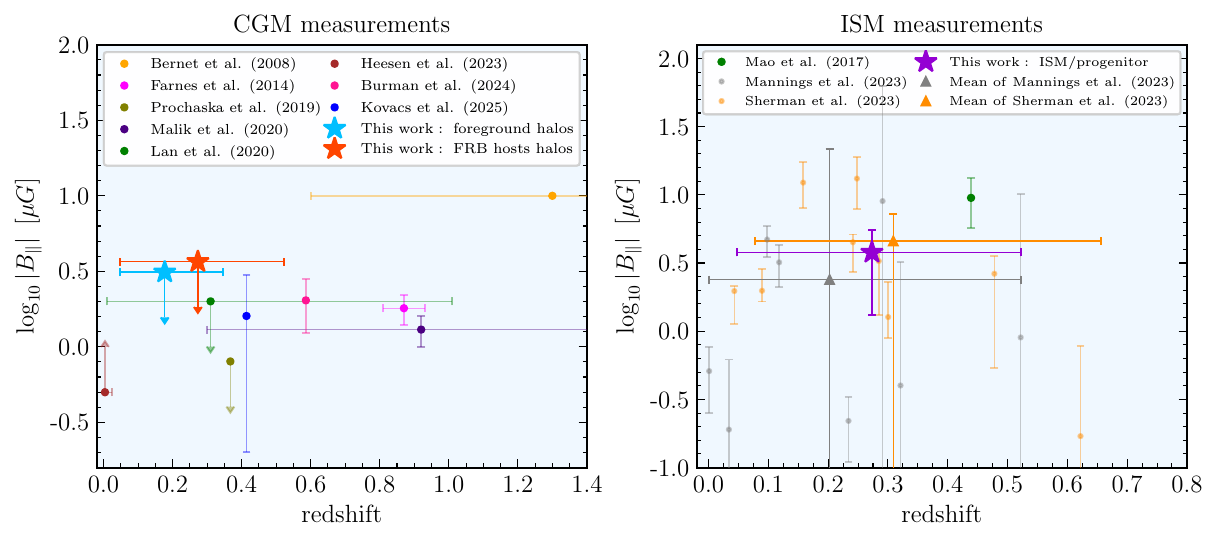}
    \caption{Comparison of $B_{\parallel}$ measurements in CGM halos obtained (left panel) and in galactic ISM (right panel), in this (star symbols) and various previous (circles) works. We place \bhosts\ and \blocal\ estimates at the average redshift of the analysed FRB sample (see Table~\ref{tab:mtab}), whereas the \bhalos\ is placed at the mean redshift of the sample of foreground halos. The horizontal error bars represent the redshift ranges of the corresponding samples.}
    \label{fig:Bz}
\end{figure*}

In addition, a recent analysis by \citet{zzhang2025} provides an independent estimate of the cosmic baryon fraction in the CGM using the \texttt{CROCODILE} cosmological simulation \citep{oku2024} with varying feedback models. They evaluated \figm\ and \fcgm\ through two complementary approaches. 
First, by directly measuring the CGM mass within halos in their fiducial run, adopting 
$r_{\mathrm{max}}= r_{200}$ and a halo mass range of  $10.5 \lesssim \log_{10}(M_{\mathrm{halo}}/M_\odot) \lesssim 15.5$, they report $\fcgm \approx 0.11$–$0.16$ at $z\simeq 0.28$. This total can be broken down into $f_{\mathrm{cgm}}\approx 0.07-0.08$ from halos with $\log_{10}(M_{\mathrm{halo}}/M_\odot) = [10.5, 13.1]$ and  $f_{\mathrm{cgm}}\approx 0.09$ from halos with $\log_{10}(M_{\mathrm{halo}}/M_\odot) = [13.1, 15.5]$, corresponding to $f_{\rm cgm}^{\rm sample} \approx 0.07-0.08$ and $f_{\rm cgm}^{\rm other} \approx 0.09$ in our notation. Notably, despite their lower number density, the more massive halos contribute disproportionately to $f_{\mathrm{cgm}}$, owing to their larger gas reservoirs. Second, they derive a semi-analytic estimate of the statistically averaged intersection fraction using projected density profiles from the simulations. Applying their Eq.~(24) yields $f_{\rm cgm}^{\rm sample}\approx 0.06-0.09$ from halos with  
$\log_{10}(M_{\mathrm{halo}}/M_\odot) = [10.5, 13.1]$, 
and again $f_{\rm cgm}^{\rm other} \approx 0.07-0.08$ in halos with $\log_{10}(M_{\mathrm{halo}}/M_\odot) = [13.1, 15.5]$.  

These simulation-based values are somewhat smaller than, though consistent within uncertainties with, our fiducial inference of $f^{\mathrm{sample}}_{\mathrm{cgm}}=0.14^{+0.07}_{-0.06}$ for 
$10 \lesssim \log_{10}(M_{\mathrm{halo}}/M_\odot) \lesssim 13.1$
at $\langle z \rangle\simeq 0.27$. (see Section~\ref{sec:results}). 
This correspondence between two independent approaches --one with forward modelling of the RM using foreground halo catalogues, and the other direct integration of simulation density profiles-- provides a valuable cross-check and highlights the utility of such comparisons for constraining feedback models in the future. 

\subsection{Magnetic fields in galactic environments at $0.1 \lesssim z \lesssim 1.0$}
\label{sec:disc_mfield}

From our study of \nfrb~FRBs, we found that the average strength of the total magnetic fields in the CGM and ISM/FRB progenitor is $B \lesssim 6.5~\mu G$. In what follows, we want to place these estimate in the context of previous extragalactic measurements in the late-time Universe. To do this, we converted our results to  measurements of $B_{\parallel}$ by drawing samples from the MCMC marginalised posterior PDFs and randomly chose the orientation angle of the magnetic field vectors from the corresponding uniform distributions:

\begin{gather}
\label{eq:bpar}
\begin{aligned}
    &B_{\parallel,i} = B_{i} \cos\alpha_{i}~,\\ 
    &B_{i} = \{ B_{i,1},~...,~B_{i,N} \} \sim p_{\rm mcmc}\left( \Theta | \rmobs \right)~,\\ 
    &\alpha_{i} = \{ \alpha_{i,1},~...,~\alpha_{i,N} \} \overset{i.i.d.}{\sim}  \mathcal{U}\left[ 0,\pi \right]~,
\end{aligned}
\end{gather}
where $i$ index denotes a given environment probed by our sample (i.e. foreground halos, halos, and local ISM environment of the FRB hosts). Applying Eq.~(\ref{eq:bpar}) to the MCMC chains yields $\approx 4000$ realisations of $B_{\parallel}$ for each of the magnetic fields we considered in this work. We estimated $B_{\rm host,\parallel}^{\rm local} = 3.8_{-2.5}^{+1.7}~\mu G$, $B_{\rm host,\parallel}^{\rm halo} \lesssim 3.7~\mu G$, and $B_{\rm fg,\parallel}^{\rm halo} \lesssim 3.1~\mu G$, respectively. Because our inference yielded constraints on the magnetic field strength associated both with the CGM of galaxies and the ISM (and/or FRB progenitors), we compared our results with those of the previous works based on a variety of techniques separately. 

The majority of the CGM constraints come from the analysis of radio-polarisation measurements towards high-$z$ quasars that present strong intervening \ion{Mg}{II} absorbers \citep[associated with galactic halos; e.g.][]{bernet2008, farnes2014, malik2020, burman2024}. Another approach is to correlate the observed \rmobs\ with the distribution of the foreground galaxies \citep[e.g.][]{lan2020, heesen2023}. An alternative method to estimate \bmag~is to use FRB sight lines that happen to pass through individual known foreground halos \citep[pioneered by][]{x2019} or via the lensing of a source's polarised emission in the ISM/halo of a foreground galaxy \citep{mao2017, kovacs2025}. Figure~\ref{fig:Bz} presents a compilation of these measurements. It is apparent from the left panel of Figure~\ref{fig:Bz} that our measurements are in good agreement with previous estimates. We note that \citet{x2019} reported an upper limit that is marginally consistent with our measurement. Moreover, we conclude that foreground or FRB host halos might contribute (on average) a non-negligible amount of RM (thus far usually neglected) and must be taken into account when analysing future observed \rmobs~of the FRBs. 

In the right panel of Figure~\ref{fig:Bz}, we compare our \blocal~estimate to previous constraints on the magnetic field in the ISM of FRB hosts \citep{mannings2023, sherman2023} and galaxy discs \citep{mao2017}. The grey circles show the results obtained by \citet{mannings2023} from the analysis of eight FRB sight lines. Similarly, the orange circles show the \bmag~measurements from the subsample of nine \texttt{DSA}-110 FRBs from \citet{sherman2023}. In addition to plotting the individual measurements, we also calculated and report the average \bmag~and a corresponding uncertainty for these two samples, and we illustrate the resulting quantities via the grey and orange triangle markers, respectively. When calculating the average of the \citet{mannings2023} sample, we excluded the \bmag~estimate for FRB~$20121102$A ($|\bmag| \simeq 777\pm366~\mu G$) as a clear outlier, most likely dominated by the immediate environment to the FRB. The average redshift values of \citet{mannings2023} and \citet{sherman2023} samples are $\langle z \rangle = 0.2023$ and $\langle z \rangle = 0.2739$, respectively, whereas ours is $\langle z \rangle = 0.2732$. We therefore artificially shifted the corresponding orange triangle by $\Delta z = 0.035$ in the figure for the sake of clarity. Finally, we also show the results of the RM analysis of the individual lensed galaxy from \citet{mao2017}. Upon inspection of the right panel of Figure~\ref{fig:Bz}, it is apparent that we find \bmag~values comparable with the average estimates from previous works. We note, however, that both of the previous works that analysed FRB RMs did not have information about the foreground halos and attributed all the \rmeg~to the contribution from the local environment in the FRB host galaxies. The fact that \bhalos, \bhosts, and \blocal\ are all of the same order of magnitude implies that all these components must be taken into account in the \rmobs. 

We remind the reader that our inferences are by construction average $B$ (or average \bmag), and thus deviations from these are expected for individual cases. Based on this reasoning, a possible explanation for the apparent peak in \bmag\ at $z\sim 0.2$ (see Figure~\ref{fig:rm_vs_z}) seen in the combined samples of \citet{mannings2023} and \citet{sherman2023} could most likely be attributed to individual excesses of \blocal\ rather than \bhosts\, given that much larger density variations are expected (due to geometrical effects). Moreover, within the \blocal~contributions of the ISM and/or local environment of the FRBs, we deem the latter to be the most likely to be responsible for outliers (including the possible peak at $z\sim 0.2$). This hypothesis can be tested with an analysis of magnetic fields from larger samples of FRBs. Additional constraints on \blocal~will also be paramount for establishing the origin of the FRB phenomenon.

Finally, we note that our analysis is agnostic with regard to magnetic field reversals in different galactic environments traversed by the FRB sight lines. While this effectively assumes the uniformity of magnetic fields at different galactic scales, given the current limited data sample we opted not to explicitly model it. A more realistic scenario of entangled fields would result in a factor of $\sqrt{N}$ amplification of the inferred $B$ values, where $N$ is the number of reversals in a given medium. We leave a more refined consideration of magnetic field correlation lengths and field reversals to future work.

\section{Conclusions}
\label{sec:end}

We analysed a sample of RMs from $N_{\rm frb}=\nfrb$~FRBs in the $0.05 \lesssim z_{\rm frb} \lesssim 0.5$ redshift range and correlated it with spectroscopic information about the foreground galactic halos, intersected by the FRB sight lines. We developed a novel Bayesian inference formalism that allows the inferring of the total strength of the magnetic fields present in various galactic environments traversed by the FRB sight lines. The main results of our work are summarised as listed below.

\begin{enumerate}
  \item For the first time, we successfully disentangled and measured the magnetic field both in the halos and ISM/progenitor environment of the FRB host galaxies. Given our fiducial set of priors, we estimate that, on average, the strength of the magnetic field in the FRB host ISM and/or the FRB progenitor is $\blocal = 5.4^{+1.1}_{-0.9}~\mu{\rm G}$, whereas we place an upper limit on the magnetic field associated with the FRB host halos $\bhosts \lesssim 4.8~\mu{\rm G}$. 

  \item In addition, we estimate an upper limit on the magnetic field in the halos of the foreground galaxies and groups; our analysis yielded $\bhalos \lesssim 4.3~\mu{\rm G}$, which is comparable to the value estimated for the halos of the FRB host galaxies. We note that for a current dataset we cannot exclude the possibility of $\bhalos = 0~\mu G $ or $\bhosts = 0~\mu G $ on a $2\sigma$ level.
  
  \item We find that our constraints are in good agreement with previous estimates of the strength of the parallel component of the magnetic field in the CGM of galaxies, as well as in the ISM environment. We stress that future attempts to measure the magnetic field using FRBs must consider the contribution from the foreground halos. 

  \item Adopting the mNFW model to describe the radial density profile of the halos of both foreground and FRB host galaxies, we estimated the average fraction of halo baryons inside the ionised CGM of individual galaxies to be $\fgas = 0.45^{+0.21}_{-0.19}$. This corresponds to the fraction of cosmic baryons $\fcgm = 0.14^{+0.07}_{-0.06}$ inside the $10.0 \lesssim \log_{10}\left( M_{\rm halo} / M_{\odot} \right) \lesssim 13.1$ halos.

  \item Our estimates for magnetic fields in the halos of foreground and FRB host galaxies, as well as in the host ISM and/or FRB progenitor environment, are largely unaffected by the average \figm~value adopted in the \fdiffz~prior (see Section~\ref{sec:priros}). On the other hand, the constraint on \fgas~is degenerate with the \figm~measurement (see discussion in Section~\ref{sec:disc_figm}).

\end{enumerate}

The results of this work emphasise the significance of the spectroscopic information regarding the foreground structures traversed by the FRB for placing accurate constraints on both cosmological and astrophysical parameters. Multiplex instruments such as \texttt{MUSE} \citep[][]{bacon2010}, \texttt{4MOST} \citep{dejong2019}, and \texttt{DESI} \citep{levi2013} will play a crucial role in interpreting DM and RM information from a surging number of FRB detections owing to increasing capabilities of the \texttt{ASKAP/CRAFT} \citep{macquart2010,shannon2025}, \texttt{CHIME/VLBI} \citep{andrew2025}, \texttt{CHIME/FRB Outriggers} \citep{lanman2024, chime2025}, and MeerKAT TRAnsients and Pulsars \citep[\texttt{MeerTrap;}][]{sanidas2018,rajwade2024}, as well as, \texttt{DSA}-110 \citep{ravi2023} and commissioning of the \texttt{DSA}-2000 \citep{hallinan2019}. These new data will allow the probing and characterisation of the properties of the magnetic fields in and out of galaxies up to redshift $z\sim1$ and beyond.

\begin{acknowledgements}
We are grateful to the anonymous referee for comments and suggestions, which helped us to improve the manuscript. We thank Joscha Jahns-Schindler and members of the Fast and Fortunate for FRB Follow-up collaboration for useful discussion and comments. I.S.K. and N.T. acknowledge support from grant ANID / FONDO ALMA 2024 / 31240053. N.T. acknowledges support by FONDECYT grant 1252229. KN acknowledge support from MEXT/JSPS KAKENHI Grant Numbers JP19H05810, JP20H00180, and JP22K21349.
Kavli IPMU was established by World Premier International Research Center Initiatives (WPI), MEXT, Japan. 
This work was performed in part at the Center
for Data-Driven Discovery, Kavli IPMU (WPI),
\end{acknowledgements}

\bibliography{aa57213-25}
\bibliographystyle{aa}

\begin{appendix}

\section{Inference test on the mock datasets}
\label{sec:appendix}
\ctable[
caption = {Inference test results for different combinations of the input true model parameters $\Theta$. },
label = {tab:itest},
width = 1.0\columnwidth,
pos = t, center]{lcccccc}
{
}
{                   \FL
 ID & \fgas & \bhalos & \bhosts & \blocal & $p\left( \in 68\%\right)$ & $p\left(\in 95\%\right)$  \NN 
    &       & $\mu G$ & $\mu G$ & $\mu G$ & &  \ML
 M01 & 0.4 & 2.0 & 3.0 & 6.0 & $75\%$ & $97\%$  \NN
 M02 & 0.7 & 2.0 & 3.0 & 6.0 & $62\%$ & $97\%$  \NN
 \FL
 M03 & 0.4 & 0.0 & 5.0 & 10.0 & $77\%$ & $96\%$  \NN
 M04 & 0.4 & 3.0 & 2.0 & 6.0 & $76\%$ & $95\%$  
\LL
}

In order to test the accuracy of our inference algorithm described in Section~\ref{sec:infernece}, we apply it to the mock FRB sample. In what follows, we briefly outline the algorithm to generate mock FRB samples and the inference results.

First, we randomly draw redshifts for $N=\nfrb$ mock FRB host galaxies from the corresponding uniform distribution $\mathcal{U}\left[ z_{\rm min}, z_{\rm max}\right]$, where the redshift range is defined to be similar to that of our dataset in Table~\ref{tab:mtab}). Second, for each mock FRB galaxy in the sample we assign halo masses by sampling the corresponding uniform distribution $\mathcal{U}\left[ M_{\rm halo}^{\rm min}, M_{\rm halo}^{\rm max}\right]$, where the halo mass range is given by the minimum and maximum halo masses in the dataset. Finally, for each mock FRB we create catalogues of the foreground halos. We begin by assigning the number of foreground halos per sightline. To do so we sample the uniform distribution $\mathcal{U}\left[ N_{\rm fg}^{\rm min}, N_{\rm fg}^{\rm max} \right]$, where $N_{\rm fg}^{\rm min}$ and $N_{\rm fg}^{\rm max}$ are given by the number of foreground halos per sightline in the dataset (see Figure~\ref{fig:fg_distr}). Then, each $i$th foreground halo along the $j$th FRB sightline is assigned a redshift $z_{i}^{j}$, impact parameter from the FRB sightline $b_{i}^{j}$, and halo mass $M_{{\rm halo},i}^{j}$

\begin{gather}
\label{eq:bparap}
\begin{aligned} 
    &\{z^j\} = \{ z_{i=1}^{j}, .., z_{i=N_{\rm fg}}^{j} \} \overset{i.i.d.}{\sim}  \mathcal{U}\left[ 0, z_{\rm FRB}^{j} \right]~,\\ 
    &\{b^j\} = \{ b_{i=1}^{j}, .., b_{i=N_{\rm fg}}^{j} \} \overset{i.i.d.}{\sim}  \mathcal{U}\left[ 30, 3000 \right]~{\rm kpc}~,\\ 
    &\{M_{\rm halos}^j\} = \{ M_{\rm halo,{i=1}}^{j}, .., M_{\rm halo,i=N_{\rm fg}}^{j} \} \overset{i.i.d.}{\sim}  \mathcal{U}\left[ 10^{10}, 10^{13.1} \right]~M_{\odot}~,\\
\end{aligned}
\end{gather}
where the boundaries of the intervals on which the uniform distributions are defined, are chosen to represent the observed dataset. We then assume a set of true model parameters $\Theta \equiv \{ \fgas=0.7; \bhalos=2.0~\mu{\rm G}; \bhosts=3.0~\mu{\rm G}; \blocal=6.0~\mu{\rm G}; \}$, and estimate the corresponding mock \rmobs~given by eq.~(\ref{eq:rmmodel}) for each mock FRB sightline. We also include an observational uncertainty by randomly sampling the Gaussian with $\sigma = 2~{\rm rad~m^{-2}}$ and adding this error to the resulting mock \rmobs values.

\begin{figure}
    \centering
    \includegraphics[width=\columnwidth]{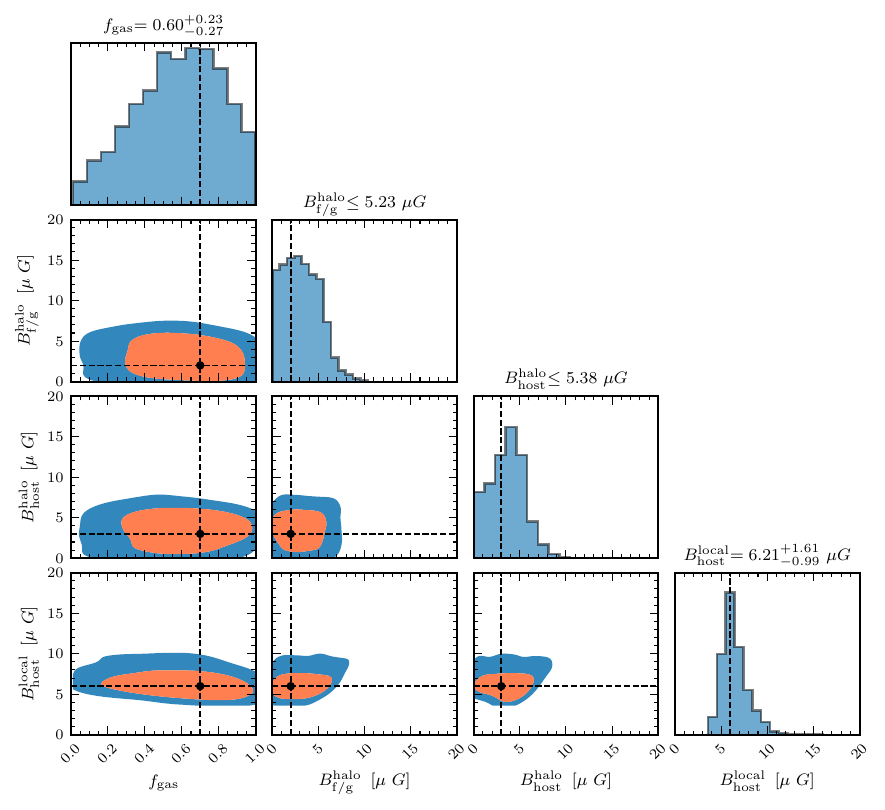}
    \caption{MCMC inference result for one realization of the mock sample of \nfrb~FRBs in M02 model (see Table~\ref{tab:itest}). The red (blue) contours correspond to the inferred $68\%$ ($95\%$) confidence intervals. The diagonal panels show the corresponding marginalized $1$D posterior probabilities of the model parameters. The black dots with dashed lines illustrate the input true values of the model parameters.}
    \label{fig:mcmc_itest}
\end{figure}

Utilizing the catalogues of the foreground halos, and \rmobs~of each mock FRB in the sample, we then proceed to calculate the joint likelihood of the mock sample, following the discussion in Section~\ref{sec:like}, and sample it with the MCMC algorithm. The resulting posterior PDFs are illustrated in Figure~\ref{fig:mcmc_itest}, where the orange (blue) contours correspond to the $68\%$ ($95\%$) confidence intervals (CI), whereas the marginalized posterior PDFs for each of the model parameters are shown in the diagonal panels. The true values of the model parameters are illustrated by the dots in each panel of Figure~\ref{fig:mcmc_itest}. Similar to the discussion in Section~\ref{sec:results}, if the maximum of a given 1D marginalized posterior PDF is at least four times larger than the larger of the two posterior PDF values at the edges of the corresponding parameter range, then we quote the $50{\rm th}$ percentile of the marginalized posterior distributions as the measured values, while the corresponding uncertainties are estimated from the $16{\rm th}$ and $84{\rm th}$ percentiles, respectively. On the other hand, if the above criterion is not met, we report an upper limit by quoting the $84$th percentile of the posterior PDF for a given parameter.

It is apparent from the inference results in Figure~\ref{fig:mcmc_itest}, that our algorithm successfully recovers the input true values of the model parameters. However, to further verify the robustness of this result, we perform an inference test. We repeat the analysis for $N=100$ different random realizations of the mock FRB sample, keeping $\Theta$ the same. Then, we calculate how often the true values of the model parameters fall inside the $68\%$ and $95\%$ CIs of the inferred posterior distributions, i.e., $p\left( \in 68\%\right)$ and $p\left( \in95\%\right)$, respectively. For the analysis to be robust, the fraction of realizations inside each of the CI should be close to the probability level of the corresponding CI. We find $p\left(\in 68\%\right) = 62\%$ and $p\left(\in 95\%\right) = 97\%$, respectively. Therefore, because these probabilities are so close to the nominal probabilities of the corresponding CIs, we conclude that our algorithm is indeed robust. Moreover, we test different realizations of $\Theta$ as well, finding similar probability values. The results are summarized in Table~\ref{tab:itest}. 

\end{appendix}
\end{document}